\documentstyle[12pt]{article}
\begin{document}
\baselineskip2pc
\title{On the Casimir of the group $ISL(n,R)$ and its algebraic decomposition}
\author{J.N. Pecina-Cruz\footnotemark\ \\Bureau of Economic
Geology\\J.J. Pickle Research Campus\\
The University of Texas at Austin\\University Station, Box X\\
Austin, Texas 78713-8294}
\addtocounter{footnote}{1}\footnotetext{Present address:\\Institute
for Advanced Technology, The University of Texas at Austin\\Ausin,
Texas 78712}
\date{Feb. 28, 2005}
\maketitle
\hyphenation{isomor-phism isomor-phic}
\hyphenation{sat-isfy def-f-ni-tion com-plete}
\hyphenation{de-ri-va-tion}
\hyphenation{pos-si-ble be-cause func-tions}
\hyphenation{ex-ist sy-mul-ta-ne-ous sys-tem to-tal dif-fer-en-tial
equa-tions}
\begin{abstract}
In this paper, an explicit expression for the Casimir operator (or
the Casimir invariant) of the inhomogeneous group $ISL(n,R)$ in its
enveloping algebra is proposed, which using contractions of the
tensorial indices of the generating operators $P^{\rho}$ and
$E_{\mu}^{\nu}$ may be presented in the following (slightly more
comprehensible as equation (1)) form. The Casimir is obtained by
symmetrizing this expression.  This tensor form is useful in the
classification of particles in affine gravitational gauge theories;
such as that based on $ISL(4,R)$. It is also proven that the Casimir
of $ISL(n,R)$ can be decomposed in terms of the Casimirs of its
little groups, a key point in the posterior construction of its
irreducible representations.
\end{abstract}

{\bf1. Introduction}\\
The special affine group $ISL(n,R)$ is the semidirect product of the
Abelian group of translations in $n$ dimensions and the special
linear, $SL(n,R).$ The $ISL(4,R)$ group has been chosen as the gauge
group in gauge theories of gravity \cite{hehl}; therefore, the
knowledge of its Casimirs will be necessary not only to investigate
the irreducible representations of this group, but also to provide
these theories with a wave equation.\cite{barut} It is has been
conjectured that a gauge theory for quantum gravity can be developed
by enlarging the Poincare group to the group $IGL(4,R).$ However,
the lack of invariants of this group\cite{demichev}, prevents to
classify the elementary  particles of a theory based on that gauge
group. Therefore, $ISL(4,R)$ is the best
selection\cite{hehl}\cite{sijacki}\cite{smolin}. The group
$ISL(n,R)$ has a subgroup, the Poincare group, from which stems its
importance in physics. The eigenvalues of the Casimir of the group
$ISL(n,R)$ provide quantum numbers to classify the particles of
these theories in the same way that the eigenvalues of the Casimirs
of the Poincare groups allow us to classify the particles according
to their mass and spin.  The eigenvalues of the Casimir of
$ISL(n,R)$ label the irreducible representations of the group. The
invariants are also useful ingredients in the decomposition of
reducible representations into irreducible ones. In the case of
gauge theories of gravity based on $ISL(n,R),$ it is important to
decompose the unitary irreducible representations of the group
$ISL(n,R)$ into the unitary irreducible representations of the
Poincare subgroup. This would bring a physical insight into the
behavior of the elementary particles of these theories. In section
2, we construct the formula for the Casimir of $ISL(n,R)$. In
section 3, we discuss the induction proof used to guarantee the
general validation of the formula for the Casimir of $ISL(n,R)$.
Finally in section 4, the algebraic decomposition of $ISL(n,R)$ is
achieved.\\ \\
{\bf2. Construction of the Formula for the Casimir of $ISL(n,R)$}\\
In Ref. 3 it is proved that the group $ISL(n,R)$ has one invariant.
And in Ref. 4 it is proved that the order of this invariant is
$\frac{1}{2}n(n+1)$. Based on this proof, the standard procedure for
constructing invariants by contracting tensorial indices with the
Levi Civita antisymmetric pseudo tensor and the generators of the
Lie group \cite{mirman}, we found a formula for the invariant of
$ISL(n,R).$ This expression is
given by\\ \\
$CasimirISL(n,R)=\{ \zeta_{\xi_{0}\alpha_{1},\ldots,\alpha_{n-1}\beta_{1}[
\beta_{2}(\gamma_{11})],\ldots,[\beta_{n-1}(\gamma_{n-2,1},\ldots,\gamma_{n-2,
n-2})]}^{\rho_{1}[(\theta_{11})\rho_{2}]\,\ldots,[(\theta_{n-2,1},\ldots,
\theta_{n-2,n-2})\rho_{n-1}]}P^{\xi_{0}}P^{\alpha_{1}}$
\begin{equation}
\cdots P^{\alpha_{n-1}}E_{\rho_{1}}^{\beta_{1}}E_{[(\theta_{11})\rho_{2}]}^{[
\beta_{2}(\gamma_{11})]} \cdots E_{[(\theta_{n-2,1},\ldots,\theta_{n-2,n-2})
\rho_{n-1}]}^{[\beta_{n-1}(\gamma_{n-2,1},\ldots,\gamma_{n-2,n-2})]}\}_{
symmetrized}
\end{equation}\\
\mbox{\scriptsize$\xi_{0},\theta_{ij},\gamma_{ij},\alpha_{l},\rho_{k},
\beta_{m} = 0,1,\ldots,n-1$\hspace{.6cm} $i,j=1,2,\ldots,n-2$\hspace{.6cm}$l,
k,m=1,2,\ldots,n-1$} \\ \\
where\\ \\
$E_{[(\theta_{11})\rho_{2}]}^{[\beta_{2}(\gamma_{11})]}=E_{\theta_{11}}^{
\beta _{2}}E_{\rho_{2}}^{\gamma_{11}}$\\
and\\
$E_{[(\theta_{n-2,1},\ldots,\theta_{n-2,n-2})\rho_{n-1}]}^{[\beta_{n-1}(\gamma_
{n-2,1},\ldots,\gamma_{n-2,n-2})]}=E_{\theta_{n-2,1}}^{\beta_{n-1}}E_{\theta_{
n-2,2}}^{\gamma_{n-2, 1}} \cdots
E_{\theta_{n-2,n-2}}^{\gamma_{n-2,n-3}}E_{\rho_{n-1}}^{\gamma_{n-2,
n-2}},$\\ \\
where $\zeta_{\ldots\ldots}^{\ldots}$ is given by the expression\\ \\
$\zeta_{\ldots\ldots}^{\ldots}=\epsilon_{\xi_{0}\beta_{1}\cdots\beta_{n-1}}
(\delta_{\rho_{1}\alpha_{1}}\cdots\delta_{\rho_{n-1}\alpha_{n-1}})((\delta_{
\theta_{11}\gamma_{11}})(\delta_{\theta_{21}\gamma_{21}}\delta_{\theta_{22}
\gamma_{22}})\cdots$\\
\begin{equation}
(\delta_{\theta_{n-2,1}\gamma_{n-2,1}}\cdots\delta_{\theta_{n-2,n-2}\gamma_{
n-2,n-2}})).
\end{equation}\\
The $E_{\alpha}^{\beta}$ are the generators of the general linear
group $GL(n,R)$ in the Weyl basis\cite{barut} and the $P^{\alpha}$
are the generators of the Abelian subgroup of $ISL(n,R)$. In formula
1, the following substitution must be carried out
\begin{equation}
{\bf
E}_{\alpha}^{\alpha}=E_{\alpha}^{\alpha}-E_{\alpha+1}^{\alpha+1}.
\end{equation}
The commutation relations of $ISL(n,R),$ are given then
\begin{equation}
[P^{\rho},P^{\nu}]=0\qquad [P^{\rho},E_{\mu}^{\nu}]=\delta
_{\mu\rho}P^{\nu} \qquad [E_{\mu}^{\nu},E_{\lambda}^{\tau}]=\delta
_{\nu \lambda}E_{\mu}^{\tau}- \delta _{\mu
\tau}E_{\lambda}^{\nu}\qquad
\end{equation}
The $E_{\alpha}^{\alpha},$ the generators of the general linear
group $GL(n,R),$ are substituted by the traceless generators ${\bf
E}_{ \alpha}^{\alpha}$ of the special linear group
$SL(n,R).$ In equation (2), we define $\epsilon_{\xi_{0}}=1$.\\ \\
{\bf3. The Induction Proof}\\
In Refs. 6, it is proved that the
invariant of $ISL(n,R)$ can be obtained by solving a system of
linear first order partial differential equations (LFPDE). The
system of (LFPDE) can trivially be solved for $n = 1$. Thus, the
first part of the induction is proven. In order to prove the second
part of the induction method, we assume the
formula is valid for $n=k$ and then prove that it is valid for $n=k+1.$\\
We can construct a scalar of the order required by Lemma 1, in Ref. 4, to
be the invariant of $ISL(k+1,R).$ According to this Lemma, the order of the
invariant in the generators of this group should be $\frac{1}{2}(k+1)(k+2).$
In this same reference, it is proven that the invariant for this group
would be of $k+1$ order in the generators of the translations. Therefore
the invariant for $ISL(k+1,R)$ must be of $k+1$ order in the translations
and of $\frac{1}{2}k(k+1)$ order in the non-translations generators of
the group $ISL(k+1,R).$ That is, the invariant of $ISL(k+1,R)$ must be
different from the invariant for $ISL(k,R)$ by a factor given by\\
\begin{equation}
P^{\alpha_{k}}E_{[(\theta_{k-1,1},\ldots,\theta_{k-1,k-1})\rho_{k}]}^{[\beta_{
k}(\gamma_{k-1,1},\ldots,\gamma_{k-1,k-1})]}.
\end{equation}
If we take into account the form of the invariants of $ISL(n,R)$ for n=1,2,3,4,\cite{pecina}
the formula we are assuming valid for $n=k$ and the factor given above,
we can construct a scalar given by\\ \\
$CasimirISL(k+1,R)=$ \\ \\$ \{
\zeta_{\xi_{0}\alpha_{1},\ldots,\alpha_{k-1}\alpha_{k}\beta_{1}[\beta_{2}(
\gamma_{11})],\ldots,[\beta_{k-1}(\gamma_{k-2,1},\ldots,\gamma_{k-2,k-2})][
\beta_{k}(\gamma_{k-1,1},\ldots,\gamma_{k-1,k-1})]}^{\rho_{1}[(\theta_{11})
\rho_{2}],\ldots,[(\theta_{k-2,1},\ldots,\theta_{k-2,k-2})\rho_{k-1}][(
\theta_{k-1,1,\ldots,\theta_{k-1,k-1})\rho_{k}]}}P^{\xi_{0}}P^{\alpha_{1}}
\cdots P^{\alpha_{k-1}}P^{\alpha_{k}}$\\
\begin{equation}
E_{\rho_{1}}^{\beta_{1}}E_{[(\theta_{11})\rho_{2}]}^{[\beta_{2}(\gamma_{11})]}
\cdots E_{[(\theta_{k-2,1},\ldots,\theta_{k-2,k-2})\rho_{k-1}]}^{[\beta_{k-1}(
\gamma_{k-2,1},\ldots,\gamma_{k-2,k-2})]}E_{[(\theta_{k-1,1},\ldots,\theta_{
k-1,k-1})\rho_{k}]}^{[\beta_{k}(\gamma_{k-1,1},\ldots,\gamma_{k-1,k-1})]}\}_{
symmetrized}.
\end{equation}\\ \\
This formula coincides with the eqn. (1) for $n=k+1.$ However, the
proof is not yet complete, since the scalar that we have constructed
to be the invariant of $ISL(k+1,R)$ could be zero. Therefore, we
must prove that the scalar given by eqn. (5) is not zero.\\ \\
{\bf4. The Algebraic Decomposition of $ISL(N,R)$}\\
The proof that the scalar given by eqn. (5) does not vanish is based
on an algebraic decomposition of the Casimir of $ISL(n,R)$ in terms
of the Casimirs of its little groups. This decomposition allows an
immediate classification of the existent particles
in a theory based on $ISL(n_{1},R)$, with $n_{1}$ any number.\cite{smolin}\\
Then making all the translations equal zero except $P^{0}$ in eqn. (5).
That is,\\ \\
$\xi_{0}=\alpha_{1}=\ldots=\alpha_{k}=0,$\\ \\
therefore\\ \\$\rho_{1}=\rho_{2}=\ldots=\rho_{k}=0.$\\ \\
Hence, the Casimir with all the translations zero except $P^{0}$ is
given by\\ \\
$CasimirISL(k+1,R)=$ \\ \\
$\{
(P^{0})^{k+1}\zeta_{00,\ldots,0\beta_{1}[\beta_{2}(\gamma_{11})],\ldots,[
\beta_{k}(\gamma_{k-1,1},\ldots,\gamma_{k-1,k-1})]}^{0[(\theta_{11})0],\ldots,
[(\theta_{k-1,1},\ldots,\theta_{k-1,k-1})0]}E_{0}^{\beta{1}}E_{0}^{
\gamma_{11}}\cdots E_{0}^{\gamma_{k-1,k-1}}$\\ \\
\begin{equation}
E_{\theta_{11}}^{\beta_{2}}E_{[(\theta_{21})\theta_{22}]}^{[\beta_{3}(\gamma_{
21})]}\cdots E_{[(\theta_{k-1,1},\ldots,\theta_{k-1,k-2})\theta_{k-1,k-1}]}^{
[\beta_{k}(\gamma_{k-1,1},\ldots,\gamma_{k-1,k-2})]}\}_{symmetrized}
\end{equation}
where\\ \\
$\zeta_{00,\ldots,0\beta_{1}[\beta_{2}(\gamma_{11})],\ldots,[\beta_{k}(
\gamma_{k-1,1},\ldots,\gamma_{k-1,k-1})]}^{0[(\theta_{11})0],\ldots,[(
\theta_{k-1,1},\ldots,\theta_{k-1,k-1})0]}=\epsilon_{0\beta_{1}\cdots
\beta_{k}}(\delta_{\theta_{11}\gamma_{11}}\delta_{\theta_{22}\gamma_{22}}
\cdots\delta_{\theta_{k-1,k-1}\gamma_{k-1,k-1}})$\\ \\
\begin{equation}
((\delta_{\theta_{21}\gamma_{21}})(\delta_{\theta_{31}\gamma_{31}}\delta_{
\theta_{32}\gamma_{32}})\cdots(\delta_{\theta_{k-1,1}\gamma_{k-1,1}}\cdots
\delta_{\theta_{k-1,k-2}\gamma_{k-1,k-2}})).
\end{equation}\\ \\
The terms of eqn. (6) with the $\theta's=0$ generated by the contraction
of the Levi-Civita pseudo tensor cancel out by antisymmetry.\\
Therefore, the indices  $\beta, \theta, \gamma$ can be shifted; instead  of
running from $0,1,\ldots,k$ they will run from $0,1,\ldots,k-1$. {\em This
defines an isomorphism} between the subset of the generators, belonging to
the factor which multiply $P^{k+1}$ in eqn. (6), of the Lie algebra of
$ISL(k+1,R)$ and the Lie algebra of $ISL(k,R).$\\
Hence eqn. (7) can be given by\\ \\
$\zeta_{\beta_{1}\gamma_{11}\gamma_{22},\ldots,\gamma_{k-1,k-1}\beta_{2}[
\beta_{3}(\gamma_{21})],\ldots,[\beta_{k}(\gamma_{k-1,1},\ldots,\gamma_{
k-1,k-2})]}^{\theta_{11}[(\theta_{21})\theta_{22}],\ldots,[(\theta_{k-1,1},
\ldots,\theta_{k-1,k-2})\theta_{k-1,k-1}]}=\epsilon_{\beta_{1}\cdots\beta_{k}}
(\delta_{\theta_{11}\gamma_{11}}\delta_{\theta_{22}\gamma_{22}}\cdots\delta_{
\theta_{k-1,k-1}\gamma_{k-1,k-1}})$\\ \\
\begin{equation}
((\delta_{\theta_{21}\gamma_{21}})(\delta_{\theta_{31}\gamma_{31}}\delta_{
\theta_{32}\gamma_{32}})\cdots(\delta_{\theta_{k-1,1}\gamma_{k-1,1}}\cdots
\delta_{\theta_{k-1,k-2}\gamma_{k-1,k-2}})).
\end{equation}\\
The basis elements of the Lie algebra of the group $ISL(n,R)$ can be
represented by the $n+1$ by $n+1$ matrices given below:
\[ \left( \begin{array}{cc}
SL(n,R)&P\\ \\
      0&0\\  \\
\end{array}  \right) \]\\
where $P$ are the generators of the group of translations, and $SL(n,R)$ are
the generators of the special linear group in $n$ dimensions. Therefore,
the $E_{0}^{\alpha}$ generators of the eqn. (6) can
be considered as the translations $P^{\alpha}$ generators of
the little group $ISL(k,R)$ of $(p^{0},0,0,\ldots,0_{k}).$\cite{elliott}\\
Using eqn. (8), eqn. (6) can be written in the following form:\\ \\
$CasimirISL(k+1,R)=$\\ \\
$\{(P^{0})^{k+1}\zeta_{\beta_{1}\gamma_{11}\gamma_{22},\ldots,\gamma_{k-1,k-1}
\beta_{2}[\beta_{3}(\gamma_{21})],\ldots,[\beta_{k}(\gamma_{k-1,1},\ldots,
\gamma_{k-1,k-2})]}^{\theta_{11}[(\theta_{21})\theta_{22}],\ldots,[(\theta_{
k-1,1},\ldots,\theta_{k-1,k-2})\theta_{k-1,k-1}]}E_{0}^{\beta{1}}E_{0}^{
\gamma_{11}}\cdots E_{0}^{\gamma_{k-1,k-1}}$\\ \\
\begin{equation}
E_{\theta_{11}}^{\beta_{2}}E_{[(\theta_{21})\theta_{22}]}^{[\beta_{3}(
\gamma_{21})]}\cdots E_{[(\theta_{k-1,1},\ldots,\theta_{k-1,k-2})\theta_{
k-1,k-1}]}^{[\beta_{k}(\gamma_{k-1,1},\ldots,\gamma_{k-1,k-2})]}\}_{
symmetrized}.
\end{equation}\\
To take advantage of the isomorphism given above, we make the following
substitution:\\ \\
$\beta_{1}\rightarrow\xi_{0}, \beta_{2}\rightarrow\beta_{1}, \beta_{3}
\rightarrow\beta_{2},\ldots,\beta_{k}\rightarrow\beta_{k-1}$\\ \\$\theta_{11}
\rightarrow\rho_{1}, \theta_{22}\rightarrow\rho_{2},\ldots,\theta_{k-1,k-1}
\rightarrow\rho_{k-1}$\\ \\$\gamma_{11}\rightarrow\alpha_{1},\gamma_{22}
\rightarrow\alpha_{2},\ldots,\gamma_{k-1,k-1}\rightarrow\alpha_{k-1}$\\ \\
$\theta_{21}\rightarrow\theta_{11}, (\theta_{31}\rightarrow\theta_{21},
\theta_{32}\rightarrow\theta_{22}),\ldots,(\theta_{k-1,1}\rightarrow\theta_{
k-2,1}, \theta_{k-1,2}\rightarrow\theta_{k-2,2},\ldots,\theta_{k-1,k-2}
\rightarrow\\ \theta_{k-2,k-2})$\\ \\$\gamma_{21}\rightarrow\gamma_{11}, (
\gamma_{31}\rightarrow\gamma_{21}, \gamma_{32}\rightarrow\gamma_{22}),\ldots,(
\gamma_{k-1,1}\rightarrow\gamma_{k-2,1}, \gamma_{k-1,2}\rightarrow\gamma_{
k-2,2},\ldots,\gamma_{k-1,k-2}\rightarrow\\ \gamma_{k-2,k-2})$\\ \\
and by substituting into eqn. (9), we obtain\\
\begin{equation}
CasimirISL(k+1,R)=\{(P^{0})^{k+1}(CasimirISL(k,R))\}_{symmetrized}
\end{equation}\\
We have obtained the Casimir of the little group $ISL(k,R)$ from the Casimir of
the group $ISL(k+1,R).$ We arrive at the same result if we take any of the
other translations.\\
From the above discussion, it is clear that the Casimir of $ISL(k+1,R)$
given by equation (5) does not vanish, as claimed. This completes the
induction proof. We conclude that the formula  given by equation (1) is
valid for any integer $n.$\\ \\
{\bf Conclusion}\\
Although the formula for the Casimir of
$ISL(n,R)$ has been written in the Weyl basis, this does not limit
its application range. The advantage of our formula for $ISL(n,R),$
over other possible formulation, is its immediate physical and
mathematical application as shown above in the little group Casimir
decomposition of $ISL(n,R).$\\
In gauge theories of gravity based on the group $ISL(4,R),$ it
should be verified the correct usage of the Casimir operator. The
reason is that in these theories, the group $ISO(1,3)$ must be a
subgroup of the gauge group. This group has a different Lie algebra
than that of the group $ISO(4)$ which is a subgroup of $ISL(4,R)$.
The applications of a deunitarizing inner
automorphism,\cite{sijacki} which changes some of the generators of
the group $ISL(n,R)$ by a factor  $\sqrt -1$, is necessary to extend
the range of application of our formula. To avoid confusion we
suggest using the notation
$ISL(1,n-1,R)$ for the group that has as a  subgroup $ISO(1,n-1).$\\ \\
{\bf Acknowledgments}\\
 I am indebted to Dr. Bob Hardage for his
constant encouragement during the preparation of this manuscript. I
thank Prof. Greg Plaxton, from the Computer Sciences Department of
The University of Texas at Austin, for many useful dicussions and
suggestions. I am also grateful to Prof. L. Wolfenstein and Prof. F.
Gildman for their support at Carnegie Mellon University while this
article was concluded.\\

\end{document}